\documentclass[preprint,preprintnumbers,amssymb,prd,nofootinbib,tightenlines]{revtex4}
\usepackage{graphicx}
\usepackage{bm}
\newcommand{\non}{\nonumber}

\begin{document}
     
\baselineskip=15pt
\parskip=5pt
   
\preprint{UK/TP-2008-03} 

\vspace*{0.7in}  

\title{\large\bf CP Asymmetries in ${\mathbf B\to f_0 K_S}$ Decays}

\author{Rupak Dutta${}^{1}$}
\email{rdutta@uky.edu}   
\author{Susan Gardner${}^{2}$}   
\email{gardner@pa.uky.edu}
\affiliation{
${}^1{}^,{}^2$Department of Physics and Astronomy, University of Kentucky, 
Lexington, Kentucky 40506-0055
\vspace{1ex} \\
}

\begin{abstract}
We consider the branching ratio and
the CP asymmetries in $B\to f_0(980) K_S $ decay  
to the end of determining the deviation of
the time-dependent CP asymmetry from $\sin(2\beta)$, $\Delta S_{f_0 K_S}
\equiv -\eta_{f_0K_S} S_{f_0K_S}-\sin(2\beta)$, 
arising from Standard Model physics. 
We obtain $\Delta S_{f_0 K_S}$ within the context of the
QCD factorization framework for the $B\to f_0(980) K_S$ decay
amplitudes assuming the $f_0(980)$ is a $q\bar{q}$ state and 
employing a random scan
over the theoretical parameter space to assess the possible range in $\Delta
S_{f_0\,K_S}$. Imposing the value of the experimental branching ratio 
within 1$\sigma$ and 3$\sigma$, respectively, of
its central value as a constraint, 
we find the range of 
$\Delta S_{f_0 K_S}$ to be $[0.018, 0.033]$  
for a scan in which the parameters are allowed to vary within 1$\sigma$
of their central values 
and the range $[-0.019, 0.064]$ 
for a scan in which the parameters vary 
within 3$\sigma$ of their central values. 
\end{abstract}

\maketitle

\section{Introduction}   
\label{intro}

In the Standard Model (SM), all CP-violating
effects derive from a single, complex phase of the 
Cabibbo-Kobayashi-Maskawa (CKM) matrix
and predicate a distinctive pattern of CP-violation~\cite{nirquinn}. 
For example, in the decay of a B-meson to a CP-eigenstate $f$, 
the time-dependent asymmetry $S_f$ realized from $b \to s c\bar c$ decay 
determines $\sin(2\beta)$, where $\beta$
is given by ${\beta} = {\arg(-V_{cd}^{} V_{cb}^\ast/V_{td}^{}
V_{tb}^\ast)}$ and $V_{ij}$ is a CKM matrix element~\cite{Carter,Bigi,Dunietz}. 
This quark-level transition can 
be studied in a variety of B-meson decays, and 
departures of the determined time-dependent asymmetry from $\sin(2\beta)$
could signal the presence of non-SM physics, which may occur in $B-\bar B$
mixing, in the decay amplitude, or in both~\cite{newphys}. 

In this paper, we consider the decay $B\to f_0(980) K_S$, which is 
mediated by the $b\to s q\bar q$ transition at one-loop-order 
in the weak interaction. 
The decay $B\to f_0(980) K_S$ is one of several penguin-dominated modes
which probe $\sin(2\beta)$. 
In contrast, in $B\to J/\psi K_S$ decay, and related charmonium modes, 
the $b\to s c\bar c$ transition operates at tree level. 
Were the time-dependent asymmetries in tree- and
penguin-dominated modes to differ, then non-SM physics could be at work
in the penguin process~\cite{grossworah}. Current experimental results 
suggest that this could be the case~\cite{talk}, 
though definite conclusions require
both experimental results of improved precision and theoretical estimates
of the subleading SM corrections. 
The numerical size of the SM corrections depend
on the specific decay mode, 
mimicking the appearance of non-SM physics~\cite{buchhiller}, 
so that the needed estimates demand 
some care. In this context 
it is worth noting that the ``wrong phase'' penguin contribution, 
proportional to the weak phase of $b\to s u\bar u$,
is particularly small in $B\to J/\psi K_S$ decay; 
indeed, the deviation of the time-dependent asymmetry $S_{J/\psi K_S}$ 
from $\sin(2\beta)$ is ${\cal O}(10^{-3})$~\cite{mannel,Li} --- 
it is suppressed by both CKM and loop effects. Thus the comparison
of this asymmetry to a ``tree-only'' determination of $\sin(2\beta)$
permits a sensitive assay of new physics in $B^0-\bar B^0$ mixing. 
Currently this last is consistent with $S_{J/\psi K_S}$, as well as with 
other determinations of $\sin(2\beta)$ which employ 
information on the sides of the unitarity triangle, 
at the ${\cal O}(10\%)$ level~\cite{ckmfitter,utfit}.  
In the case of the penguin modes, the wrong-phase
penguin is larger as it is suffers only CKM, i.e., 
${\cal O}(\lambda^2)\simeq 0.04$, suppression. 
In these modes the computed SM deviations from  $\sin(2\beta)$ determine a
much-needed baseline against which the experimental results 
can be assessed for new-physics effects, as 
new-physics-induced deviations from
sin$2\beta$ could certainly be channel-dependent as well. 
Systematic studies of the SM 
corrections exist~\cite{glq,beneke,soni} in a variety of modes. 
The $B\to f_0(980) K_S$ mode has 
received less attention, perhaps due to the 
ill-known quark structure of the $f_0(980)$~\cite{cheng,ck1}; 
this is a deficiency we wish to remedy. 

The $f_0(980)$ is a fairly narrow resonance of non-Breit-Wigner
form which couples to $\pi\pi$ and $K \bar K$ final states. 
The quark structure of the $f_0(980)$ meson is not well established.
Much discussion has revolved around whether it is better regarded
as a $q^2 \bar q^2$ state~\cite{jaffe} or, perhaps, as a $K\bar K$ 
molecule~\cite{weinisgur,Lohse}. The theoretical correction 
relevant to the interpretation of $S_{f_0\,K_S}$ as a measurement 
of sin$2\beta$ comes from the presence of the 
$b\to su \bar u $ transition in the decay amplitude, and 
interconnected issues complicate its analysis. Not only 
must we consider the possible non-$q\bar q$ structure of 
the $f_0$ resonance, but we must also recall that 
strong final-state interactions exist in $\pi\pi$ scattering in the $I=0, J=0$
channel~\cite{Truong,Dobado,Gasser}. Such final-state interactions 
give rise to the scalar form factor, and particularly the
partial width $\Gamma (f_0(980) \to \pi\pi\,,K\bar K)$, 
which can be computed through the unitarization of a scattering kernel 
compatible with 
low-energy constraints~\cite{olleroset,ollerulf,lahdeulf}. 
We presume the $f_0(980)$ resonance to be sufficiently narrow relative
to the energy released in the $B$ decay 
that we 
can approximate the full decay amplitude as the product of the
two-body decay amplitude $A(B\to f_0(980) K_S)$ 
with $\Gamma (f_0(980) \to \pi\pi\,,K\bar K)$, where we refer to 
Ref.~\cite{svgulf} for a discussion of the assumptions implicit
to this treatment. In particular, the strong phases associated with
the long-distance physics of  
$f_0$ decay are presumed to be universal and not modified by
the B-decay environment, so that the decay-specific phases are
captured by the application of 
QCD perturbation theory in 
the heavy quark limit to the two-body decay process. 
Corrections to the picture we employ can be estimated, 
though not in a systematically improvable way. 
For example, so-called final-state 
rescattering has been estimated for various $b\to s$ penguin 
modes~\cite{soni}, and OZI-violating
effects can also be considered, the latter contributing
significantly to $J/\psi \to \phi \pi\pi$ decay, e.g.~\cite{ollerulf,lahdeulf}. 
We reserve discussion of the impact of 
the computed 
$f_0\to \pi\pi$ and $f_0 \to K\,\bar{K}$ partial widths, as well as of
possible OZI-violating effects, to a subsequent publication~\cite{etal}. 
Our particular focus is in the study of the sensitivity of
our prediction of the CP asymmetries to 
the assumed quark structure of the $f_0(980)$; the latter 
enters in the evaluation of the hadronic matrix elements of the
$b\to s u \bar u$ transition. A priori one might think 
such effects to be most 
important in the assessment of the deviation of $S_{f_0K_s}$ from 
$\sin(2\beta)$. We proceed in the same vein as Cheng and
collaborators~\cite{cheng,ck1},
treating the $f_0(980)$ as a $q\bar q$ state and 
employing the 
QCD factorization approach~\cite{bbns,BN}  
for the hadronic matrix elements. We differ from this earlier work in
the treatment of the $B \to f_0$ form factor.
We then perform a random scan over
the theoretical parameter space, 
after Beneke~\cite{beneke}, 
to assess, in part, the sensitivity of the results to the employed
hadronic matrix elements and hence to the implicitly assumed
structure of the $f_0(980)$. 

Let us give a brief outline of this paper. 
We begin, in Sec.~\ref{TB}, with a description of the theoretical framework
and briefly review the role of the $b\to s q\bar q$ transition, 
with $q\in u,d$, in the determination of the 
CP-violating 
parameters $S_f$  and $C_f$, as well as of the two-body
branching ratio. We present a synopsis of the pertinent 
QCD factorization formulae as well. 
In 
Sec.~\ref{RND} we present the input parameters we use in our
numerical calculations, reporting the CP-violating
parameters which result from the use of our defined 
``default'' set
of input parameters. It should be emphasized that the factorization theorem
from which the QCD factorization approach follows holds only at leading
power in the heavy quark mass, and the estimate of 
$1/m_b$ power suppressed terms, though of apparent phenomenological 
importance, is uncertain. 
Thus our random scan over the theoretical
parameter space, effected to explore the possible range of $\Delta S_{f_0\,K_S}$ 
and $C_{f_0\,K_S}$, takes both uncertainties in the theoretical inputs and
in the assessment of the ${\cal O}(\Lambda_{QCD}/M_B)$ corrections 
into account. We report these results in Sec.~\ref{RND} as well. 
We conclude with a summary of our results and an outlook on future
work in Sec.~\ref{cn}.

\section{Theoretical Framework}
\label{TB}

We pattern our analysis after that of Beneke, Ref.~\cite{beneke}, 
and, indeed, adopt a common notation. 
In particular, we employ the QCD factorization framework~\cite{bbns,BN} 
for the computation of the $B\to f_0 K_S$ matrix elements~\cite{cheng} 
and perform a random scan over 
the space of possible input parameters to estimate the 
uncertainty in their computation. 
We begin by recalling that the 
CP asymmetry $A_f(t)$ into a CP eigenstate $f$
is given by 
\begin{equation}
A_{f}(t) = \frac{\mbox{Br}(\bar B^0(t)\to f) - \mbox{Br}(B^0(t)\to
f)}{\mbox{Br}(\bar B^0(t)\to f) + \mbox{Br}(B^0(t)\to f)} 
\equiv  S_f\sin(\Delta M_B\,t) - C_f\cos(\Delta M_B\,t)  \,.
\end{equation}
Here we have neglected $\Delta \Gamma$ where $\Delta \Gamma \equiv
\Gamma_H - \Gamma_L$ is the width difference of the $B$ eigenstates.
We note $\Delta M_B \equiv M_H - M_L$ 
is the mass difference of the 
$B$
eigenstates, $S_f$ is the CP asymmetry generated by the
interference of $B$ - $\bar{B}$
mixing and direct decay, and $C_f$ is an asymmetry reflective of 
direct CP violation. We recall~\cite{nirquinn} 
\begin{equation}
S_f={2\,{\rm Im}\bm{\lambda}_f\over 1+|\bm{\lambda}_f|^2} \quad; \quad
C_f={1-|\bm{\lambda}_f|^2\over 1+|\bm{\lambda}_f|^2} \,,
\end{equation}
where
\begin{equation}
\bm{\lambda}_f=\left(\frac{q}{p}\right)_B\,\frac{A(\bar B^0 \to f)}{A(B^0\to f)}\,, 
\end{equation}
with the factor $(q/p)_B$ characterizing $B-\bar B$ mixing. If we treat the $f_0(980)$ resonance as
if it were a stable particle, as in Ref.~\cite{cheng}, then $f\equiv f_0(980) K_S$,
and we write 
\begin{equation}
\bm{\lambda}_{f_0 K_S}=\left(\frac{q}{p}\right)_B\,\left(\frac{q}{p}\right)_K\,
\frac{A(\bar B^0 \to f_0 \bar K^0)}{A(B^0\to f_0 K^0)}\,, 
\end{equation}
where the factor $(q/p)_K$ characterizes $K-\bar K$ mixing. 
In this event, $f$ is a two-body final state, and 
we write the decay amplitude as 
\begin{equation}
A(\bar B\to f) = \lambda_c \,a_{f}^c + 
\lambda_u\,a_{f}^u
\propto 
(1 + e^{-i\gamma} \,d_{f})\,, 
\end{equation}
with $d_{f} \equiv |{\lambda_u}/{\lambda_c}| 
({a_{f}^u}/{a_{f}^c})$ 
and $\lambda_q\equiv V_{qb} V_{qs}^\ast$ 
for $q\in u,c$  
to determine
\begin{equation}
\label{dSS}
\Delta S_f \equiv -\eta_f S_f-\sin(2\beta) =
\frac{2 \,\mbox{Re}(d_f) \cos(2\beta)\sin\gamma +|d_f|^2
\left(\sin(2\beta+2\gamma)-\sin(2\beta)\right)}
{1 + 2 \,\mbox{Re}(d_f) \cos\gamma + |d_f|^2}\,, 
\end{equation}
where $\eta_f$ is the eigenvalue of the CP-operator 
associated with the eigenstate $f$.
Moreover, we have 
\begin{equation}
C_f = - \frac{2 \,\mbox{Im} (d_f)
\sin\gamma}
{1 + 2 \,\mbox{Re} (d_f) \cos\gamma + |d_f|^2} \,,
\label{CC}
\end{equation}
so that if $|d_f|$ is small, the functions $\Delta S_f$ and $C_f$ 
show little correlation~\cite{beneke}. 
The value of 
$\Delta S_f$ in the SM shows certain systematic trends with $f$~\cite{beneke}. 
For example, 
if a color-suppressed tree amplitude $C$ contributes to $a_{f}^u$, then 
 $\Delta S_f$ is much larger than it would be if it were absent. 
In the
latter event, the $u$-quark penguin amplitude $P^u$ drives $a_{f}^u$. 
Generally, $d_f \propto (\pm C + P^u)/P^c$. If the parameters describing
the $B\to f$ decay amplitude make $|P^c|$ 
small relative to $|C|$, then $\Delta S_f$ can range over a 
wide array of values. Such large excursions can be controlled, however, 
by demanding that the amplitudes be consistent 
with the empirical branching ratios~\cite{beneke}. 
In the case of current interest, $B\to f_0 K_S$ decay, 
there is no $C$ amplitude, so that we expect $\Delta S_{f_0 K_S}$ 
to be small on general grounds, and the imposition of a branching
ratio constraint should no longer be crucial. 
However, we have found exceptional regions in 
theoretical parameter space 
for which $P^c$ is small with respect to $P^u$ and thus $\Delta S_{f_0\,K_S}$
is large, so that 
it is,  in fact, crucial to apply the branching ratio constraint to
eliminate these large excursions. 

If we treat $\bar B^0 \to f_0(980) \bar K^0$ 
decay as a two-body process, then the
decay rate is 
\begin{equation}
\label{G}
\Gamma = \frac{p}{8\,\pi\,M_B^2} |\mathcal{M}|^2, 
\end{equation}
where $\mathcal{M}\equiv A(\bar B^0 \to f_0 \bar K^0)$, 
as we shall describe in detail, and 
\begin{equation}
p=\frac{\sqrt{(M_B^2-(M_{f_{0}}+M_{K^0})^2)(M_B^2-(M_{f_{0}}-M_{K^0})^2)}}{2M_B} \,. 
\end{equation}
We recall that the branching ratio is given by 
${\Gamma}/{\Gamma_B}$, where $\Gamma_B$ is the total decay width of $B$ meson. 
In reality, the $f_0(980)$ is not a stable particle; rather, it 
is a resonance which decays 
to both $\pi\pi$ and $K\bar K$ final states. 
We wish to investigate the role of finite-width effects explicitly in
a subsequent publication~\cite{etal}: we neglect them here. 
To begin, we rewrite the two-body decay amplitude 
as 
\begin{equation}
A(\bar B \to f_0 \bar K_0) = A_u^n \lambda_u + A_c^n \lambda_c + 
A_u^s \lambda_u + A_c^s \lambda_c \,,
\label{decomp}
\end{equation} 
where the superscript $(n,s)$ refers to the non-strange and strange
quark components of the $f_0(980)$, respectively, so that 
\begin{equation}
d_{f_0\,K_S} = \Big|\frac{\lambda_u}{\lambda_c}\Big|\left(\frac{A_u^n + A_u^s}{A_c^n +
A_c^s}\right)\,.
\label{df}
\end{equation}
We now proceed to calculate the $A_{u,c}^{n,s}$ amplitudes.

\subsection{${\mathbf  {\bar B}\to f_0 {\bar K}}$ Decay in QCD Factorization}
\label{QCD}
The decay amplitudes of exclusive hadronic $B$-meson decays can be 
systematically analyzed in a combined expansion of inverse powers of the 
heavy quark mass $m_b$ and the strong coupling constant
$\alpha_s$~\cite{scet1,scet2,scet3}. 
The QCD factorization approach, in specific, permits the rigorous computation of
these amplitudes, for certain two-body final states, 
in leading power in ${\cal O}(\Lambda_{QCD}/m_b)$
and in a power series  
in $\alpha_s(\mu)$, where 
$\mu \sim {\cal O}(m_b)$~\cite{bbns}. 
Its starting point is the effective
weak Hamiltonian for charmless hadronic $B$ decay, consisting of a sum of
the products of CKM matrix elements, local operators $Q_i$, 
and Wilson coefficients $C_i(\mu)$, 
evaluated in next-to-leading order (NLO) 
precision in $\alpha_s(\mu)$~\cite{bucha}. 
Some dissension in the literature exists 
concerning the ingredients of the leading
power analysis~\cite{hnli}, particularly in regards to the charm-quark
penguin contributions~\cite{Bauer:2004tj,Beneke:2004bn}:  
in the QCD factorization 
approach any non-factorizable charm-quark penguin contributions appear 
in ${\cal O}(\Lambda_{QCD}/m_b)$ corrections. Although recent work suggests
that such charm-quark penguin effects may be needed to 
explain the $B$-decay data
to light charmless mesons~\cite{Jain:2007dy}, note also 
Ref.~\cite{charming}, 
in our context we can safely neglect such
developments, as they cannot make $\Delta S_{f_0\,K_S}$ larger. 
$B$-meson decays to scalar- and pseudoscalar-meson final states 
have recently been studied by Cheng and collaborators in a series of 
papers~\cite{ck1,cheng,soni}; they 
employ the QCD factorization approach and treat the $f_0(980)$ resonance 
as a $q\bar q$ state. Their results connect to those 
of Beneke and Neubert for $B$-meson decays 
to pseudoscalar- and vector-meson final states~\cite{BN}, as 
the amplitudes for the scalar- and pseudoscalar-meson 
channels follow from these earlier results upon a series 
of replacements~\cite{cheng}. This means that the twist-3 light-front
distribution amplitudes are assumed to be determined by 
two-particle configurations only. 
The amplitude for 
$\bar B^0 \to f_0 \bar K^0$ decay is given in Ref.~\cite{cheng}, so 
that we identify, after Eq.~(\ref{decomp}), 
\begin{eqnarray}
\label{eq:A}
A_u^n &=& - \frac{G_F}{\sqrt{2}} \left[\left(a_4^u-r_\chi^Ka_6^u
-{1\over 2}(a_{10}^u-r_\chi^K a_8^u)\right)_{f_0 K}\right]
f_KF_0^{Bf_0}(M_{K^0}^2)(M_B^2-M_{f_0}^2) 
\non \\
&&+
\frac{G_F}{\sqrt{2}}
\left[\left( b_3 -{1\over
2}b_{\rm 3EW}\right)_{f_0^d K}\right]f_B\, , \\
A_c^n &=& - \frac{G_F}{\sqrt{2}} \left[\left(a_4^c-r_\chi^Ka_6^c
-{1\over 2}(a_{10}^c-r_\chi^K a_8^c)\right)_{f_0 K}\right]
f_KF_0^{Bf_0}(M_{K^0}^2)(M_B^2-M_{f_0}^2) 
\non \\
&&+
\frac{G_F}{\sqrt{2}}
\left[\left( b_3 -{1\over
2}b_{\rm 3EW}\right)_{f_0^d K}\right]f_B\, , \\
A_u^s &=& - \frac{G_F}{\sqrt{2}} \left[\left(a_6^u
-{1\over 2}a_{8}^u\right)_{K f_0}\right]
\bar r_\chi^{f_0} \bar f_{f_0}^s F_0^{BK}(M_{f_0}^2)(M_B^2-M_{K^0}^2) 
\non \\
&&+
\frac{G_F}{\sqrt{2}}
\left[\left( b_3 -{1\over
2}b_{\rm 3EW}\right)_{K f_0^s}\right]f_B\, , \\
A_c^s &=& - \frac{G_F}{\sqrt{2}} \left[\left(a_6^c
-{1\over 2}a_{8}^c\right)_{K f_0}\right]
\bar r_\chi^{f_0} \bar f_{f_0}^s F_0^{BK}(M_{f_0}^2)(M_B^2-M_{K^0}^2) 
\non \\
&&+
\frac{G_F}{\sqrt{2}}
\left[\left( b_3 -{1\over
2}b_{\rm 3EW}\right)_{K f_0}\right]f_B\,. 
\end{eqnarray}
The $M_1 M_2$ subscripts mean that the quantities in brackets
are to be interpreted as $a_i^p(M_1 M_2)$, in which the $M_1$ meson 
contains the spectator quark from the $B$-meson, 
and $b_i(M_1 M_2)$, in which the $M_1$ meson carries an anti-quark 
from the weak vertex and the $M_2$ meson contains a quark from the weak vertex. 
The QCD factorization framework is rigorous 
in leading order in ${\cal O}(\Lambda_{QCD}/m_b)$, but it also 
includes estimates of $1/m_b$-suppressed corrections. We note that 
$\bar r_\chi^{f_0} = 2M_{f_0}/m_b(\mu)$ and 
$r_\chi^{K} = 2M_{K^0}^2/(m_b(\mu)(m_s(\mu) + m_d(\mu)))$ but that these 
chirally-enhanced terms are counted as terms of leading power 
in the $1/m_b$ expansion in QCD factorization. The 
quark masses are running masses defined in the $\overline{MS}$ scheme. 
Certain $1/m_b$ corrections can 
suffer endpoint divergences, and their estimate is uncertain. 
The $b_i(M_1 M_2)$ terms, for example, 
reflect annihilation contributions, which 
are a class of $1/m_b$-suppressed corrections to the decay rate. Moreover, 
the $a_i^p(M_1 M_2)$ terms contain 
${\cal O}(\alpha_s)$ corrections to the operator matrix elements,
which can, in turn, 
contain $1/m_b$-suppressed corrections. The ${\cal O}(\alpha_s)$
corrections encode 
non-perturbative input through integrals over the 
light-cone distribution amplitudes. 
The coefficients $a_i^p(M_1 M_2)$, where $p = u , c$,  can be expressed
as~\cite{bbns,BN}  
\begin{equation}
\label{aip}
 a_i^p(M_1M_2) = C_i(\mu)+{C_{i\pm1}(\mu)\over N_c}
  +{C_{i\pm1}(\mu)\over N_c}\,{C_F\alpha_s(\mu) \over
 4\pi}\Big[V_i(M_2)+{4\pi^2\over N_c}H_i(M_1M_2)\Big]+P_i^p(M_2),
 \end{equation}
where the upper signs apply when $i$ is
odd, the lower signs apply when $i$ is even, and 
$C_F=(N_c^2-1)/2N_c$ with $N_c=3$.  
The quantities $V_i(M_2)$, $H_i(M_1M_2)$, and $P_i(M_2)$ 
reflect vertex corrections, hard spectator interactions, 
and penguin corrections, respectively. The function 
$H_i(M_1M_2)$ contains a endpoint divergence in its power suppressed terms --- 
we refer to Refs.~\cite{bbns,BN} for all omitted details.
For work towards a theory of the power corrections, 
we refer the reader to the developments in Refs.~\cite{scet1,scet2,scet3}. 
The coefficients $b_i$ relevant to our calculation can be expressed
as~\cite{bbns,BN}
\begin{eqnarray}
\label{bi}
b_3(M_1M_2) = \frac{C_F}{N_c^2}\Big[C_3\,A_1^i + C_5(A_3^i + A_3^f) +
N_c\,C_6\,A_3^f\Big]\,,  \nonumber \\
b_{3,EW}(M_1M_2) = \frac{C_F}{N_c^2}\Big[C_9\,A_1^i + C_7(A_3^i + A_3^f)
+ N_c\,C_8\,A_3^i\Big]\,,
\end{eqnarray}
where in the annihilation amplitudes $A_n^{i,f}\equiv A_n^{i,f}(M_1 M_2)$ 
the superscripts $i$ and $f$ refer to gluon
emission from initial- and final-state quarks, respectively. 
For the calculation of $A_n^{i,f}(M_1 M_2)$ 
we include the 
corrections coming from $\alpha_2^K$, the second Gegenbauer moment of
the kaon, for consistency with our analysis of the $f_0(980)$, 
and we provide explicit expressions
in App.~\ref{app}. 

The expressions 
for $\bar B \to f_0 \bar K$ decay contain non-perturbative 
hadronic input through the meson decay constants and $B\to M$ form factors;
these quantities are sensitive to the assumed quark structure of
the hadrons. Assuming the $f_0(980)$ resonance can
be written as a $q\bar q$ state, we write 
\begin{equation}
|f_0(980) \rangle = \cos \theta |s\bar s \rangle + \sin \theta 
|n\bar n \rangle \,,
\label{f0mix}
\end{equation}
where $|n\bar n \rangle 
\equiv (|u\bar u \rangle  + |d\bar d \rangle )/\sqrt{2}$. 
The empirical observation of both 
$\Gamma(J/\psi \to f_0 \omega)$ and $\Gamma(J/\psi \to f_0 \phi)$ 
suggests that the $f_0(980)$ has both strange and
non-strange components, so that $\theta$ is non-zero~\cite{ck1}. 
We define the $B\to f_0$ form factor as 
\begin{equation}
\langle f_0(p^\prime)|
\bar d \gamma_\mu \gamma_5 b
|B(p)\rangle = -i\left[\left(P_\mu - \frac{M_B^2 -
M_{f_0}^2}{q^2}\,q_\mu\right)F_1^{Bf_0}(q^2) + \frac{M_B^2 -
M_{f_0}^2}{q^2}\,q_\mu\,F_0^{Bf_0}(q^2)\right] \,,
\end{equation}
where $P_\mu = (p + p^\prime)_\mu$  and  $q_\mu = (p - p^\prime)_\mu$, 
and the $f_0$ decay constant as 
\begin{equation}
\langle f_0 | q\bar q | 0 \rangle = M_{f_0} \bar f^q_{f_0} 
\end{equation}
for $q \in (n,s)$.
Defining $| f_0^q\rangle \equiv | q \bar q \rangle$, we have 
\begin{equation}
\langle f_0^s | s\bar{s} |0 \rangle = M_{f_{0}}\,\tilde{f}_{f_{0}}^s 
\quad\quad \hbox{and} 
\quad\quad
\langle f_0^n | u\bar{u} |0 \rangle = \frac{1}{\sqrt{2}}
M_{f_{0}}\,\tilde{f}_{f_{0}}^n  \,,
\end{equation}
so that 
$\bar{f}_{f_{0}}^s = \tilde{f}_{f_{0}}^s\,\cos\theta$ and 
$\bar{f}_{f_{0}}^n = \tilde{f}_{f_{0}}^n\,\sin\theta$. 
Similarly we define
$F_0^{Bf_0} = \sin\theta\,F_0^{Bf_0^d}/{\sqrt{2}}$,
where $F_0^{B f_0^d}$ describes the form factor to the 
$| d \bar d \rangle$ piece of the $f_0(980)$ final state. 

In order to calculate the $B \to f_0$ form factor we assume the $f_0$ to
be a $q\bar{q}$
state and use the constituent quark
model (CQM) of 
Refs.~\cite{Gatto,Deandrea1,Deandrea2,Polosa},
which combines
heavy quark effective theory with chiral symmetry in
the light quark sector. In Refs.~\cite{Gatto,Deandrea1}, Gatto 
{\it et al.} study the
$D \to \sigma \pi \to 3\pi$ and $D_s \to
f_0 \pi$ amplitude using the CQM model for the $D \to \sigma$ and $D_s
\to f_0$ form factors and find good agreement with E$791$ data.
This model describes
interactions in terms of effective vertices between a light quark,
a heavy quark, and a heavy meson. The model depends on both its UV 
and IR cutoffs. The UV cutoff $\Lambda$ is set by the spontaneous chiral
symmetry breaking scale $\Lambda_{\chi}$, which 
is of ${\cal O}(1\,{\rm GeV})$. This model does not include
confinement, so that
one has to introduce an IR cutoff $\bar \mu$. The constituent quark mass $m$,
is determined by solving
the Nambu Jona-Lasinio (NJL) gap equation, which, in turn,
depends on the UV and IR cutoffs. We choose $\Lambda = 1.25\, {\rm GeV}$
as described in Ref.~\cite{Polosa}. 
Thus for fixed $\Lambda$, as the IR
cutoff varies, $m$ varies
accordingly, as explicitly illustrated in 
Ref.~\cite{Ebert}. 
For the default value of the $B\to f_0$
form factor we use $m = 0.3\, {\rm GeV}$, $\bar\mu = 0.3\, {\rm GeV}$, and
$\Delta_H = 0.4\,{\rm GeV}$. The parameter 
$\Delta_H \equiv M_H - M_Q$, where
$M_H$ is the mass of the heavy meson
and $M_Q$ is the mass of the constituent heavy quark.
Explicit expressions for the polar and 
direct contributions to the form factor are given
in Refs.~\cite{Gatto,Deandrea1}. 
To assess the range of the $B\to f_0$ form factor we vary $\bar\mu$ in
the range $[0.25, 0.35]$, as per the associated variation in $m$ given in 
Ref.~\cite{Ebert}, 
 and vary $\Delta_H$ in the range $[0.3, 0.5]$ as given in 
Refs.~\cite{Gatto,Deandrea1,Deandrea2,Polosa}. 
We find that the variation in $F^{Bf_0^{d}}$ is mainly driven by the variation
in $\bar \mu$. However, 
the error range reported in Refs.~\cite{Gatto,Deandrea1} 
was apparently determined by varying $\Delta_H$ alone~\cite{pc}. 
The default value of the $B\to f_0$ form factor, as well as
its uncertainty, are given in Sec.~\ref{in}.

The uncertainty in the calculation of the decay amplitude comes from 
both the statistical and systematic errors present in the QCD
factorization approach. The uncertainties associated with different
input parameters such as the scalar meson decay constants, the form factors,
the quark masses, and the 
Gegenbauer moments of light-cone distribution amplitudes
are of first kind, whereas the uncertainties associated
with the $1/m_b$-suppressed corrections are of the second kind.
The theoretical uncertainties in the latter case come from the hard
spectator and the weak annihilation contributions which contain endpoint
divergences. The endpoint divergences $X_H$ in the hard spectator and
$X_A$ in the annihilation terms are parameterized as~\cite{bbns}
\begin{equation}
X_{H,A} =
\ln\left(\frac{M_B}{\Lambda_h}\right)(1+\rho_{H,A}e^{i\phi_{H,A}})\,,
\end{equation}
where we assume $\rho_{H,A} \le 1$ and $\Lambda_h = 0.5\, \rm{GeV}$. 
We note $\phi_{H,A}$ are unknown, strong-interaction phases.

We wish to determine the impact of the various theoretical uncertainties 
in the $B\to f_0 K$ decay amplitude 
on the value of $\Delta S_{f_0K_S}$ in a quantitative way. 
To realize this, we begin by defining  a ``default model.''
This consists of using  
the central values of the inputs given in Ref.~\cite{cheng,bbns}, as well  
as setting $\rho_A=\rho_H=0$ 
in the parameterization of the endpoint divergences. 
With this in place we thus determine the default values of the 
$A_{u,c}^{n,s}$ amplitudes.
To gauge the size of the uncertainties, we perform a 
random scan of the allowed theoretical parameter space. 
That is, we include the
uncertainties coming from all the input parameters. 
For the parameter scan, we choose the range of the input
parameters by taking either 1$\sigma$, 2$\sigma$, or 3$\sigma$ deviations from
the central values. Once we have the maximal and minimal values of
the inputs in the chosen ranges, we draw random values of the
input parameters within that range. Similarly, to include
the uncertainties coming from $X_A$ and $X_H$ we vary $\rho_{H,A}$
from $0$ to $1$ and 
$\phi_{H,A}$ from $0$ to $2\pi$. That is, we draw random values of
$\rho_{H,A}$ in the range $0$ to $1$ and $\phi_{H,A}$ in the range $0$
to $2\pi$.

\section{Results and Discussion}
\label{RND}

\subsection{Inputs}
\label{in}
For definiteness, we summarize the parameter choices of our
default model. 
We employ a renormalization scale of ${\cal O}(m_b/2)$; namely, 
we choose $\mu = 2.1\,{\rm GeV}$, so that 
the value of the strong coupling constant in NLO at this scale 
is $\alpha_s(2.1\,{\rm GeV})= 0.286$. 
To realize this, 
we work with $N_f=5$ throughout, after Ref.~\cite{bbns}, 
and choose $\Lambda^{(5)}=0.225$ GeV, which 
corresponds to $\alpha_s(M_Z)= 0.118$. As a check of
the accuracy of NLO precision in this context, we compare
our results with those computed at $\mu = 4.2\,{\rm GeV}$, 
for which we note $\alpha_s(m_b)= 0.224$.
For the electromagnetic coupling constant, 
we use $\alpha^{-1} = 129$
and neglect the $Q^2$ evolution of $\alpha$ as in Ref.~\cite{bbns}.
To realize the Wilson coefficients, 
we follow the procedures of Ref.~\cite{bbns}, using 
$m_t(m_t)=167$ GeV, $M_W=80.4$ GeV, and 
$\sin^2 \theta_W =0.23$~\cite{bbns}
\footnote{Repeating our calculation with $m_t=172.7\,{\rm GeV}$, 
as per Ref.~\cite{pdg}, and following the procedures described
in text yield $\Delta S_{f_0 K_S}=0.0269$, 
$C_{f_0 K_S}=-0.00557$, and ${\rm Br}(f_0 K_S)= 13.6\times 10^{-6}$
in place of the values reported in Eq.~(\ref{w}). We can safely
neglect this update in our parameter scans.}
and verify 
the leading-order (LO) and NLO Wilson coefficients they report
through explicit computation. We also 
use  the values $\rho_A=\rho_H=0$ to fix the default values of 
the hard spectator and  annihilation terms. We fix the 
value of the $b$ quark mass at the scale $\mu = 2.1\,{\rm GeV}$ to be 
$m_b(2.1\,{\rm GeV})=4.88\pm0.08\, {\rm GeV}$, which corresponds to
$m_b(m_b)=4.2\pm0.07\, {\rm GeV}$~\cite{pdg} using 
the two-loop expression for the running quark
mass~\cite{bucha}. The error in $m_b(2.1\,{\rm GeV})$ is calculated
using the maximum and the minimum value of $m_b$ at the $\mu = m_b$
scale. Similarly the value of $m_s (m_b) = 0.077\pm0.017\,{\rm GeV}$ 
corresponds to 
$m_s (2.1\,{\rm GeV}) = 0.090\pm0.020\,{\rm GeV}$, after Ref.~\cite{BN}. 
For the meson masses we use $M_B = 5.2795(5)\, {\rm GeV}$, 
$M_{K^0}=0.497648(22)\, {\rm GeV}$, and $M_{f_0} =0.980(10)\, {\rm GeV}$, 
 as given in Ref.~\cite{pdg}, where the uncertainty in the last
digits is indicated by the number in parentheses. Similarly, 
the mean life of $B$ meson is
$\tau_B = 1.530(9)\times 10^{-12}$ s~\cite{pdg}, and 
$G_F=1.16637(1)\times 10^{-5}\, {\rm GeV}^{-2}$~\cite{pdg}. 
The errors in these empirical quantities are unimportant for our purposes, 
so that we neglect them in our random scan. 
The CKM matrix elements, taken from~\cite{ckm} are,
\begin{eqnarray}
\label{ckm}
&&|V_{ub}| = 0.00357^{+0.00017}_{-0.00017}, \qquad\qquad
|V_{us}| = 0.22653^{+0.00075}_{-0.00077} \nonumber \\
&&|V_{cb}| = 0.0405^{+0.0032}_{-0.0029}, \qquad\qquad
|V_{cs}| = 0.97316^{+0.00018}_{-0.00018} \nonumber \\
&& |\lambda_u/\lambda_c| = 0.0205 \pm0.0019, \qquad\qquad
|\lambda_c| = 0.0394 \pm0.0031 \nonumber \\
&&\gamma = 76.8 ^{+30.4}_{-31.5}\,, \qquad\qquad
\beta = 21.5^{+1.0}_{-1.0}\,,
\end{eqnarray}
where $\gamma$ and $\beta$ are reported in degrees. We
use the uncertainties associated with the CKM matrix elements to
calculate the uncertainties in $|\lambda_c|$ and $|\lambda_u/\lambda_c|$,
assuming the errors are uncorrelated for simplicity. We note the angle 
$\gamma$ is the phase associated with $\lambda_u/\lambda_c$, namely
$\lambda_u/\lambda_c\equiv|\lambda_u/\lambda_c|\,e^{-i\,\gamma}$, and
that $\beta$ is the unitarity-triangle angle 
we defined in Sec.~\ref{intro}. In our scan over parameter space
we vary our inputs within 2$\sigma$ and 3$\sigma$ of their 
default values as well. 
In the case of the CKM parameters we use the ranges as reported
in Ref.~\cite{ckm} for $|V_{ik}|, \gamma$, and $\beta$ for 
$\pm 2\sigma$ and $\pm 3\sigma$ as appropriate; it is worth
noting, in particular, that $\gamma$ when ranged over a 3$\sigma$
variation is never negative. 

The other inputs are taken 
from Refs.~\cite{cheng,bbns,Ali} and 
are given as follows. 
All the scale-dependent quantities in the scalar sector 
are evaluated at $\mu = 2.1\, {\rm GeV}$ as per Ref.~\cite{cheng}. 
The ratio of charm quark mass to the $b$ quark mass is 
taken from Ref.~\cite{Ali}; the resulting value of the
charm quark mass encompasses the value recommended in Ref.~\cite{pdg}.
The value of the mixing angle is taken from Ref.~\cite{cheng}. 
All the other input parameters are taken from Ref.~\cite{bbns}.
\begin{eqnarray}
\label{un}
&&B_1 = -0.54 \pm0.06\,, \qquad\qquad
B_3 = 0.01 \pm0.04\,, \nonumber \\
&&\alpha_1^K = 0.3 \pm0.3\,, \qquad\qquad
\alpha_2^K = 0.1 \pm0.3\,, \nonumber \\
&&\lambda_B = 0.35 \pm0.15\,{\rm GeV}\,, \qquad\qquad
f_B = 0.20 \pm0.03\, {\rm GeV}\,, \nonumber \\
&&F_0^{BK} = 0.35 \pm0.03\,, \qquad\qquad
f_{K}=0.16\, {\rm GeV}\,, \qquad\qquad \nonumber \\
&&\bar{f}_{f_0} =  0.460 \pm0.025\,{\rm GeV}\,, \qquad\qquad
\theta = 152.5 \pm12.5^\circ\,, \qquad\qquad \nonumber \\
&&m_c/m_b = 0.27\pm0.06\,, \qquad\qquad
m_q/m_s = 0.0413 \,. 
\end{eqnarray}
We note that $m_q=(m_u + m_d)/2$ and that we neglect the error in 
$m_q/m_s$. This amounts to neglecting the error in the light-quark
mass $m_q$, since we include the error in the strange quark mass. 
This we may safely do as we employ $m_q/m_s$ in the evaluation of
$r_\chi^K$ exclusively; we note that mass differences of ${\cal O}(m_d-m_u)$
are tantamount to the inclusion of isospin-breaking effects,
which are numerically unimportant to us here. 
Explicit expressions for the leading-twist, light-cone distribution
amplitudes of the light mesons are given in 
Eq.~(\ref{gegen}), where 
we note $B_1$ and $B_3$ are the first and third Gegenbauer moments of the 
$f_0(980)$ and 
$\alpha_1^K$ and $\alpha_2^K$ are the first and second Gegenbauer moments of
the kaon. Although all the Gegenbauer moments are scale-dependent
in principle, the inclusion of such variations is beyond the accuracy
of the NLO treatment we effect here, so that we neglect such refinements. 
The first inverse moment of the $B$ meson light-cone distribution
amplitude, $\lambda_B$, is defined by
$\int_{0}^{1}\,dx\,\Phi_{B}(x)/x = {M_B}/{\lambda_B}$, where
$\Phi_B(x)$ is one of two twist-2 light-cone distribution
amplitudes of the $B$ meson~\cite{bbns,BN}. 
We term $f_B$, $F_0^{BK}$, $f_K$, and $f_{f_{0}}$ the $B$ meson decay
constant, the $B \to K$ 
form factor, the $K$ meson decay constant, 
and the scalar decay constant, respectively. The 
given value of 
$f_{f_{0}}$ is specific to the assumed $q\bar{q}$
structure of $f_0$; we recall $\theta$ is the mixing angle of
Eq.~(\ref{f0mix}). 
For our inputs,  
we note that the ratios $r_{\chi}^M$ for the $K$ meson and $f_0$ meson, are 
\begin{eqnarray}
&&r_{\chi}^K =  1.08\,,\qquad\qquad
\bar{r}_{\chi}^{f_0} = 0.402\,.
 \end{eqnarray}
Our default value of the $B\to f_0$ form 
factor is
\begin{eqnarray}
\label{fbf0}
F_0^{Bf_0^{d}} = 0.284\, ,
\end{eqnarray}
where we use the CQM model and the parameters
$m = 0.3\,{\rm GeV}$, $\bar \mu = 0.3\,{\rm
GeV}$, and $\Delta_H = 0.4\,{\rm GeV}$. Cheng et al.~\cite{cheng} 
assert that the $F^{Bf_0^d}$ form
factor should be comparable in size to 
the $B\to \pi$ form factor, and we find this to be 
consistent with our own numerical estimate. 
We calculate the error 
in the $B\to f_0$ form factor by varying $\bar \mu$
in the range $[0.25, 0.35]$ and $\Delta_H$ in the range $[0.3, 0.5]$.
We find that the variation in $F_0^{Bf_0^{d}}$ with respect to
$\Delta_H$ is small compared to the variation 
with respect to $\bar \mu$. 
That is, varying $\Delta_H$ over our chosen range makes 
$F_0^{Bf_0^{d}}$ range over the values $[0.28, 0.29]$.
Varying $\bar \mu$ over its chosen range as well, 
we find that $F_0^{Bf_0^{d}}$ ranges from $[0.23, 0.46]$.
For the random scan we use the range $[0.23, 0.46]$ for our
1$\sigma$ scan and triple its range, to yield 
$[0.0048,0.69]$ for our 3$\sigma$ scan. 
 
The coefficients  $a_i^p(f_0K)$,  $a_{6,8}^p(Kf_0)$ and $b_i$ are calculated
at the scale $\mu = 2.1\,{\rm GeV}$ using Eqs.~(\ref{aip}) and
(\ref{bi}), as well as the formulae in Ref.~\cite{bbns}, to yield  
\begin{eqnarray}
\label{eq:ai}
&&  a_4^u = -0.0261 - i0.0208, \qquad\qquad
a_4^c = -0.0340 - i0.0110, \nonumber \\
&& a_6^u = -0.0581 -i 0.0185, \qquad\qquad
a_6^c = -0.0641 -i 0.00842, \nonumber \\
&& a_8^u = (76.6 - i0.434)\times 10^{-5}, \qquad\quad
a_8^c = (76.5 - i0.259)\times 10^{-5},  \nonumber \\
&& a_{10}^u = (-172+i130  )\times 10^{-5}, \qquad\quad
a_{10}^c = (-172 + i130 )\times 10^{-5}, \nonumber \\
&& a_{6,8}^p(Kf_0) =   a_{6,8}^p(f_0K),  \nonumber \\
&& b_3(f_0\,K_S) = -0.0506, \qquad\qquad
b_3(K_S\,f_0) = 0.0264,\nonumber \\
&& b_{3,EW}(f_0\,K_S) = -0.00133, \qquad\qquad
b_{3,EW}(K_S\,f_0) = -0.000768,
\end{eqnarray}
Our formulae for $b_3$ and $b_{3EW}$ differ slightly from 
those given in Ref.~\cite{bbns,cheng}, as we detail in 
App.~\ref{app}, though the numerical differences are 
negligible. 
Now that we have all the input parameters, we can calculate the
default value of the $A_{u,c}^{n,s}$ amplitudes using 
Eq.~(\ref{eq:A}). We find
\begin{eqnarray}
&&  A^{n}_{u} = (-8.12+ i0.438)\times 10^{-8}\,{\rm GeV}, \qquad\qquad
A^{n}_{c} = (-7.67- i0.821)\times 10^{-8}\,{\rm GeV}, \nonumber \\
&&  A^{s}_{u} = (-84.7 - i24.2)\times 10^{-8}\,{\rm GeV}, \qquad\qquad
A^{s}_{c} = (-92.5- i11.0)\times 10^{-8}\,{\rm GeV},
\end{eqnarray}
If we ignore the annihilation terms from the $B\to f_0K_S$ decay
amplitude, we find 
\begin{eqnarray}
&&  A^{n}_{u} = (-12.5+ i0.438)\times 10^{-8}\,{\rm GeV}, \qquad\qquad
A^{n}_{c} = (-12.1- i0.821)\times 10^{-8}\,{\rm GeV}, \nonumber \\
&&  A^{s}_{u} = (-76.4 - i24.2)\times 10^{-8}\,{\rm GeV}, \qquad\qquad
A^{s}_{c} = (-84.3- i11.0)\times 10^{-8}\,{\rm GeV},
\end{eqnarray}
If we ignore all the $1/m_b$ power suppressed terms in 
the $B\to f_0K_S$ decay amplitude, we find 
\begin{eqnarray}
&&  A^{n}_{u} = (-12.1 + i0.438)\times 10^{-8}\,{\rm GeV}, \qquad\qquad
A^{n}_{c} = (-11.7- i0.821)\times 10^{-8}\,{\rm GeV}, \nonumber \\
&&  A^{s}_{u} = (-76.4 - i24.2)\times 10^{-8}\,{\rm GeV}, \qquad\qquad
A^{s}_{c} = (-84.3- i11.0)\times 10^{-8}\,{\rm GeV},
\end{eqnarray}
For our default parameter set, the amplitudes mediated
by the $s\bar s$ component of the $f_0(980)$ are numerically
dominant, and 
the various
contributions coming from the power-suppressed terms 
are small corrections to the leading power contributions. 
However, these power corrections suffer endpoint
divergences, and their estimate is uncertain; 
they could well affect 
the value of $\Delta S_{f_0\,K_S}$. We shall consider this possibility
in detail
in Sec.~\ref{23body}. 
We note that the dominance of the penguin contributions associated
with the strange quark 
component of the $f_0$ emerges because 
the $a_4$ and $a_6$ interfere
destructively exclusively in the decay 
to the nonstrange 
component of the $f_0$, as we can see from Eq.~(\ref{eq:A}). 
Moreover, we note that the contribution coming from the annihilation
amplitude in this case is smaller than 
the contribution coming from the penguin
amplitude, so that 
the $B\to f_0\,K_S$ decay amplitude is dominated by 
the $s\bar s$ component of the $f_0$.

\subsection{Two-body Decay Results}
\label{23body}
Using the default values of the $A_{u,c}^{n,s}$ amplitudes we can
easily calculate $d_{f_0\,K_S}$ and 
thus $\Delta S_{f_0\,K_S}$ and $C_{f_0\,K_S}$ using
Eq.~(\ref{dSS}) and Eq.~(\ref{CC}). 
We compute the values of $\Delta S_{f_0\,K_S}$,
$C_{f_0\,K_S}$, and the branching ratio (Br) 
for the two-body decay as discussed in
Sec.~\ref{TB}. 
For our default parameter set we find 
\begin{eqnarray}
\label{w}
&&\Delta S_{f_0K_S} = 0.0269, \qquad\qquad
   C_{f_0K_S} = -0.00561, \qquad\qquad
\hbox{Br}
(f_0K_S) =13.4\times 10^{-6} \,.
\end{eqnarray}
We did the same calculation
ignoring the annihilation terms,
for which we find 
\begin{eqnarray}
\label{w1}
&&\Delta S_{f_0K_S} = 0.0268, \qquad\qquad
   C_{f_0K_S} = -0.00584, \qquad\qquad
\hbox{Br}(f_0K_S) =12.4\times 10^{-6}  \,.
\end{eqnarray}
Alternatively, if we neglect 
all $1/m_b$ suppressed terms we find 
\begin{eqnarray}
\label{w2}
&&\Delta S_{f_0K_S} = 0.0268 , \qquad\qquad
   C_{f_0K_S} = -0.00587, \qquad\qquad
\hbox{Br}(f_0K_S) = 12.3\times 10^{-6} \,.
\end{eqnarray} 
We have done the same calculation at the $\mu = m_b$ 
scale, and our results are 
very similar to those given in 
Eq.~(\ref{w}), Eq.~(\ref{w1}), and Eq.~(\ref{w2}). 
That is, we
find the values of $\Delta 
S_{f_0\,K_S}$ are $0.0269$, $0.0269$, and $0.0269$, respectively, 
whereas the values of $C_{f_0\,K_S}$
are $-0.00603$, $-0.00621$, and $-0.00623$,  respectively.
However the value of the branching ratio does change, 
giving 
$10.3\times 10^{-6}$,  
$9.69\times 10^{-6}$, and  $9.63\times 10^{-6}$, respectively.
It is evident from our numerical estimates that $\Delta S_{f_0\,K_S}$ and $C_{f_0\,K_S}$
and the two body branching ratio receive little contribution from the
annihilation terms or, indeed, from any of the $1/m_b$ suppressed terms.
That is what we expect since the $B\to f_0K_S$ decay amplitude is 
driven by the $s$-quark component of the $f_0$; the 
leading power contributions to 
the $A_{u,c}^{s}$ amplitudes are much larger than the contributions
coming from the power suppressed terms, and 
$\Delta S_{f_0\,K_S}$ and $C_{f_0\,K_S}$ depend only on 
the ratio of the $A_{u,c}^{n,s}$ amplitudes $d_{f_0\,K_S}$.
However, the branching
ratio is controlled by the square of the sum of the 
amplitudes. Thus any change in 
the size of the amplitudes themselves 
will definitely change the branching ratio; this explains
why our branching ratios are more sensitive to the value of $\mu$. 
Our default value of the branching ratio is not
consistent with the experimental value of 
$(5.8\pm 0.8)\times 10^{-6}$~\cite{talk}
obtained by BaBar and Belle~\cite{Aubert2, Garmash2}, so that 
the default set of 
input parameters we employ can not explain the experimental data. 
This does not falsify our approach per se, as it is reasonable
to think that we can compute the ratio of amplitudes $d_{f_0\,K_S}$
more accurately than the $A_{u,c}^{n,s}$ amplitudes alone. 
We also find that we are able to confront the empirical branching
ratio successfully in our scan of the theoretical parameter space. 

The numerical values of $\Delta S_{f_0\,K_S}$ 
and $C_{f_0\,K_S}$ for $B\to f_0K_S$ decay are
calculated by Ref.~\cite{cheng} as well, for which they find 
\begin{eqnarray}
\label{c}
&& \Delta S_{f_{0}K_{S}} = 0.023, \qquad\qquad
C_{f_{0}K_{S}} = -0.008 \,.
\end{eqnarray}
Our default values of $\Delta S_{f_0\,K_S}$ and $C_{f_0\,K_S}$ for 
$B\to f_0K_S$ decay are similar to those in Ref.~\cite{cheng}, 
but they are not exactly same. We note that we
also differ in our computed values of 
$a_i^p(f_0K)$ and $a_{6,8}^p(Kf_0)$, though this emerges, at least
in part, because they employ a value of $\alpha_s(\mu=2.1\,{\rm GeV})$
appropriate to $N_f=4$. 

Our default values of $\Delta S_{f_0\,K_S}$ and $C_{f_0\,K_S}$ are small and that is what we
expect from the SM point of view. In the QCD factorization 
approach the
uncertainties in the calculation of the decay amplitude come from  the
input parameters as well as from the $1/m_b$ suppressed terms. 
We would like to establish the range  of $\Delta S_{f_0\,K_S}$ and $C_{f_0\,K_S}$
possible from SM physics using the 
effect of the various uncertainties in the $B\to f_0K_S$
decay amplitude. 
Thus it is necessary to include the corrections  
arising from the uncertainties
associated with all the input parameters, including those
in  the scalar meson decay
constants, the form factors, the quark masses, the CKM elements, and the
Gegenbauer moments of the light-cone distribution amplitudes. 
Since the quark structure 
of the $f_0$ is not well known, we are particularly interested
in whether the uncertainties in the $B\to f_0$ form factor and the 
$f_0$ scalar decay constant impact the value of $\Delta S_{f_0\,K_S}$ in
a significant way. 
Additional systematic 
uncertainties in our calculation 
come from the power corrections which 
contain endpoint divergences; our estimate of these is uncertain. 
As we have noted, too, 
the branching ratio we compute from 
our default parameter set 
does not confront the experimental 
branching ratio successfully, though this is not necessary
to describe $\Delta S_{f_0\,K_S}$ well. 

\begin{figure}[htbp!]
\rotatebox{270}{\includegraphics[height=12cm,width=8cm]{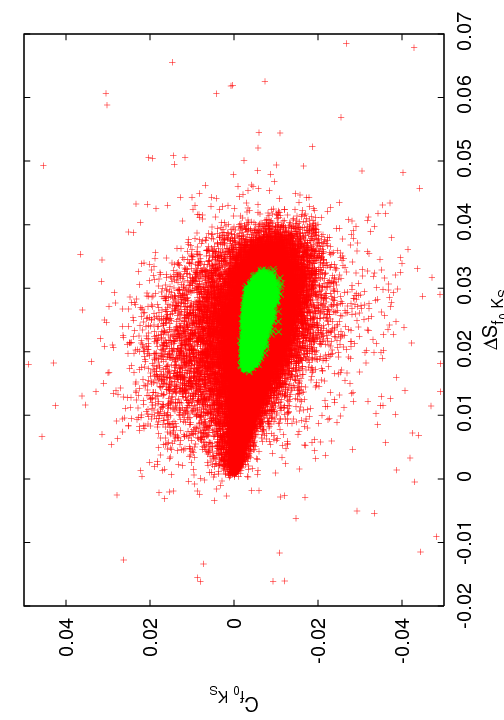}}
\caption{\label{fig1} Range in $\Delta S_{f_0\,K_S}$ and $C_{f_0\,K_S}$ 
from a scan of 500,000 theoretical models. 
The lighter interior region corresponds
to a scan of the parameter space at 1$\sigma$, where the darker, larger
region corresponds to a scan at 3$\sigma$. 
Some points in the 3$\sigma$ scan fall outside 
the window chosen for the illustration.
}
\end{figure}

To see the effect of the above mentioned uncertainties on the various
observables, 
we perform a
random scan of the allowed theoretical parameter space. The central values
of the inputs and the uncertainties associated with them are in
Eq.~(\ref{ckm}), Eq.~(\ref{un}), and Eq.~(\ref{fbf0}).
For the parameter scan, we choose the range of these
input parameters by taking 1$\sigma$ and 
3$\sigma$ variations from 
the central values. Once we have the maximal and minimal values of 
the inputs in the range mentioned above, we draw random values of the input
parameters in the chosen range. To include the uncertainties coming
from the hard spectator and the weak annihilation terms 
we vary $\rho_{H,A}$ from $0$ to $1$ and $\phi_{H,A}$ from $0$ to
$2\pi$. That is, we draw random values of $\rho_{H,A}$ in the range $0$
to $1$ and $\phi_{H,A}$ in the range $0$ to $2\pi$. We use the public
domain random number generator ``rannyu'' throughout; 
note Ref.~\cite{knuth} for a discussion of suitable inputs. 
Each chosen parameter set corresponds to a theoretical model, and 
we have plotted the various combination of
$\Delta S_{f_0\,K_S}$ and $C_{f_0\,K_S}$ for 
$500,000$ models for a particular seed, just for illustration, 
for parameter ranges fixed at 1$\sigma$ and 3$\sigma$ 
from their central values 
in Fig.~\ref{fig1}. 
The box-like shapes of the resulting regions suggest that there is 
little correlation between $\Delta S_{f_0\,K_S}$ and $C_{f_0\,K_S}$. 
That means $|d_{f_0\,K_S}|$ is
small, and $\Delta S_{f_0\,K_S}$ and $C_{f_0\,K_S}$ are mainly driven by Re($d_{f_0\,K_S}$) and
Im($d_{f_0\,K_S}$), respectively, as we can see from Eqs.~(\ref{dSS}) and (\ref{CC}).

\begin{figure}[htbp!]
\rotatebox{270}{\includegraphics[height=12cm,width=8cm]{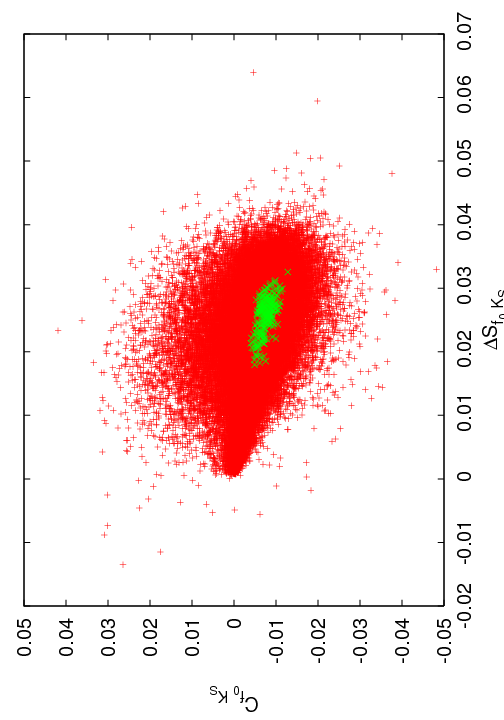}}
\caption{\label{fig2} Range in $\Delta S_{f_0\,K_S}$ and $C_{f_0\,K_S}$ from a scan of
500,000 
theoretical models with the empirical branching ratio constraint imposed. 
The darker 
points represent the possible range
of $\Delta S_{f_0\,K_S}$ and $C_{f_0\,K_S}$ within 3$\sigma$ whereas the lighter points 
represent the range for a 1$\sigma$ variation. 
} 
\end{figure}

We did a random scan for two different seeds and for each seed we
use $100,000$ and $500,000$ different theoretical 
models. We determine the range in $\Delta S_{f_0\,K_S}$
which results, or indeed of any observable, 
 by evaluating the extremal values which 
emerge from the parameter scan. This makes the results sensitive,
in principle, to the detailed manner in which the scan is effected. 
The resulting range in $\Delta S_{f_0\,K_S}$ varies little for different
values of the seed and for the different numbers of theoretical models 
if we choose the input parameters within 1$\sigma$ of the
central values. For example, the range of $\Delta S_{f_0K_S}$ for a
1$\sigma$ scan is found to be $[0.017, 0.034]$ and $[0.016, 0.035]$
for two different seeds from a sample of $500,000$ parameter sets. 
However, the range of $\Delta S_{f_0\,K_S}$ does change significantly with seed
if we choose the input parameters to range within either 2$\sigma$ or 3$\sigma$ 
of the central values. For example, the 
range of $\Delta S_{f_0K_S}$ in the scan over a 
3$\sigma$ 
variation is found to be 
$[-0.41, 0.29]$ and $[-0.34, 0.20]$ for two different seeds from a
sample of $500,000$ parameter sets. 
It is evident from our scan that though the range of
$\Delta S_{f_0\,K_S}$ is small for 1$\sigma$ variations, 
it is large for 3$\sigma$ variations. 
No color-suppressed tree amplitude contributes to 
the $B\to f_0K_S$ decay process, but we have large excursions 
in $\Delta S_{f_0\,K_S}$ over small regions of the parameter space nevertheless. 
In these special regions, the $P^c$ 
amplitude becomes small and drives 
the large values of $\Delta S_{f_0\,K_S}$. 
However, such small values of $P^c$ always
give theoretical branching ratios which are much too small. 
It thus becomes crucial to apply the 
experimental branching ratio constraint to determine 
the allowed range in $\Delta S_{f_0\,K_S}$ within the SM. 
We recall that the average branching
ratio for $B\to f_0 K_S$ decay process measured by Belle and
BaBar~\cite{Aubert2, Garmash2} is $(5.8\pm 0.8)\times 10^{-6}$. 
We impose the branching ratio constraint in such a
way that we ignore those theoretical models which are not
compatible within 1$\sigma$ of the experimental branching ratio for 1$\sigma$
parameter scans and within 3$\sigma$ of the experimental branching ratio
for the 3$\sigma$ parameter scans, respectively. 
After we impose the empirical branching ratio constraint, 
the seed-averaged range of $\Delta S_{f_0K_S}$ for our scan of models
spanning 1$\sigma$ and
3$\sigma$ variations are found to be $[0.018,0.033]$ and $[-0.019,
0.064]$, respectively, for the 500,000 point simulation. 
In this context ``seed-averaged'' means that we report the 
extremal values of $\Delta S_{f_0\,K_S}$ which result from scans using two
different seeds. 
It is interesting to note that the range in 
$\Delta S_{f_0\,K_S}$ 
once the branching ratio constraint is applied is almost identical to what
was found without the branching ratio constraint in 
the 1$\sigma$ scan. 
This underscores our point that the branching ratio
is largely independent of the value of $\Delta S_{f_0\,K_S}$. 
We find that $C_{f_0\,K_S}$ as well can range over a wide array of values giving
$[-0.47, 0.51]$ and $[-0.55, 0.49]$ for two different seeds from
a sample of
$500,000$ parameter sets within 3$\sigma$ of the default values.
However, after applying the branching ratio constraint, 
such large excursions disappear, 
giving the seed averaged range $[-0.045, 0.051]$ for 3$\sigma$ and
$[-0.013, -0.0024]$ for 1$\sigma$. 
We plot the various 
combinations of $\Delta S_{f_0\,K_S}$ and $C_{f_0\,K_S}$ with the branching ratio constraint for
the seed used in Fig.~\ref{fig1}, just for illustration, 
for a random sample of $500,000$  
models 
in Fig.~\ref{fig2}. 
After applying the branching ratio constraint, 
we find that the range in $\Delta S_{f_0\,K_S}$ and in $C_{f_0\,K_S}$ 
depends on neither the number of parameters nor the particular 
seed used in the random scan. To test the accuracy of our NLO analysis, 
we also change the renormalization
scale $\mu$ from $m_b/2$ to $m_b$, and we find very little variation in
$\Delta S_{f_0\,K_S}$ and $C_{f_0\,K_S}$ once the empirical branching ratio constraint is
imposed. That is, the range of $\Delta S_{f_0\,K_S}$ and $C_{f_0\,K_S}$ in this case 
are $[0.017, 0.032]$ and 
$[-0.012, -0.0032]$ for 1$\sigma$, respectively, 
whereas they are $[-0.015, 0.061]$ and 
$[-0.044, 0.033]$ for 3$\sigma$. The weak $\mu$ dependence we observe 
follows as the observables
involve ratios of computed amplitudes in each case. 

\begin{figure}[htbp!]
\includegraphics{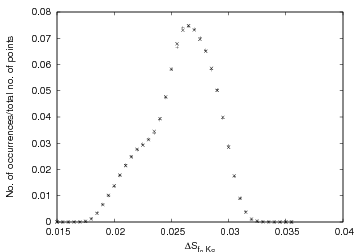}
\caption{\label{fig3} 
Histogram for $\Delta S_{f_0K_S}$ for a random
scan employing parameters which range over 1$\sigma$ of their central
values. We note $+$ denotes the scan with 100,000 models and 
$\times$ denotes the scan with 500,000 models. The range of
$\Delta S_{f_0K_S}$ is the same for each case.}
\end{figure}

\begin{figure}[htbp!]
\includegraphics{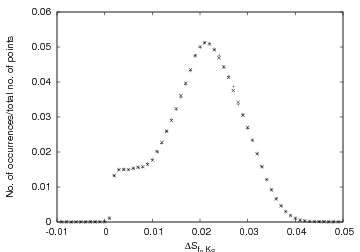}
\caption{
Histogram for $\Delta S_{f_0K_S}$ for a random
scan employing parameters which range over 3$\sigma$ of their central
values. We note $+$ denotes the scan with 100,000 models and 
$\times$ denotes the scan with 500,000 models. 
Some points in the scan fall outside the window chosen for the
illustration.
\label{fig4}}
\end{figure}

\begin{figure}[htbp!]
\includegraphics{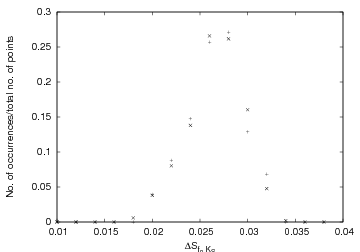}
\caption{\label{fig5} Histogram for $\Delta S_{f_0K_S}$, after
that of Fig.~\ref{fig3}, once the empirical branching ratio constraint
is imposed. 
Here the total number of points refer to the total
number of models which survive the imposed branching ratio 
constraint. 
}
\end{figure}

\begin{figure}[htbp!]
\includegraphics{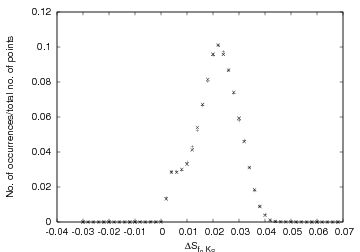}
\caption{\label{fig6}
Histogram for $\Delta S_{f_0K_S}$, after
that of Fig.~\ref{fig4}, once the empirical branching ratio constraint
is imposed. 
Here the total number of points refer to the total
number of models which survive the imposed branching ratio 
constraint. 
}
\end{figure}

To study the likelihood 
of various values of $\Delta S_{f_0K_S}$ in theory space, 
we have plotted a histogram of $\Delta S_{f_0K_S}$ 
from a sample of 100,000 and 500,000 theoretical models, after
averaging over seeds, in 
Fig.~\ref{fig3}. To plot the histogram we first set a bin size and then
determine the number of models which fall within each bin. 
Since we have different numbers of theoretical models in our random scans, 
to plot all of them together we divide the number of models falling in each bin 
by the total number of theoretical models used and 
choose that quantity as the ordinate 
of each histogram. If no points are shown, both here and in later
figures, that means that no models whatsoever occupy that region of
theory space. 
For scans employing parameters within a 1$\sigma$ variation the range 
of $\Delta S_{f_0K_S}$ varies little if we change the seed or 
the number of models used in the scan. 
However, it does vary notably if the parameters chosen range within 2$\sigma$ 
or 3$\sigma$ variations of their central values. 
In Fig.~\ref{fig4}, we plot a
histogram of $\Delta S_{f_0K_S}$, after averaging over seeds,  
from a sample of 
$100,000$ and $500,000$ models taking the values of the input
parameters within 3$\sigma$ of their central values. 
It can be seen from the histogram that although the actual range of $\Delta
S_{f_0K_S}$ varies for different seeds and for different number of parameter
sets, the shape of the histogram in 
$\Delta S_{f_0K_S}$ does not seem to vary.
We also plot the histograms of $\Delta S_{f_0\,K_S}$ after applying the
branching ratio constraint, as shown in Fig.~\ref{fig5} and
Fig.~\ref{fig6}, and note that the shape of the histogram changes little. 
These results are also seed-averaged. 
In these figures the total number of points refer to the total 
number of models which satisfy the branching ratio constraint. 
In particular, the scale of the ordinate is not same in Fig.~\ref{fig5} and
Fig.~\ref{fig6} because the total number of points which 
satisfy the branching ratio constraint at 
1$\sigma$ is small compared to those which satisfy 
the branching ratio constraint at 3$\sigma$. Note that the
integral of these model space results over $\Delta S_{f_0\,K_S}$ should
yield unity in each case. 

\begin{figure}[htbp!]
\includegraphics{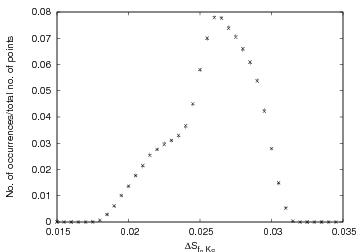}
\caption{Histogram for $\Delta S_{f_0K_S}$ varying only the CKM
parameters over a 1$\sigma$ variation. 
\label{fig7}}
\end{figure}

\begin{figure}[htbp!]
\includegraphics{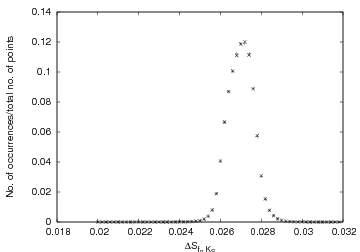}
\caption{Histogram for $\Delta S_{f_0K_S}$ varying only the hadronic
parameters over a 1$\sigma$ variation.
\label{fig8}}
\end{figure}

We have also studied the impact of the hadronic uncertainties alone on
$\Delta S_{f_0K_S}$ for the $B \to f_0K_S$ decay process. To do this we
set the CKM parameters to their default values and vary all the other
input parameters. 
We include the variations in the power corrections as
well. 
Similarly to see the impact of the CKM parameters, we
set all the other hadronic inputs to their default values and vary 
only the CKM parameters $|\lambda_c|$, $|\lambda_u/\lambda_c|$,
$\gamma$, and $\beta$. We
include the uncertainties associated with $V_{ij}$ in an uncorrelated
way to find the uncertainties in $|\lambda_c|$ and $|\lambda_u/\lambda_c|$.
We perform a
random scan over these CKM parameters within 1$\sigma$ and 3$\sigma$
of their central values as given in Eq.~(\ref{ckm}) and Ref.~\cite{ckm}.  
The associated histograms of $\Delta S_{f_0K_S}$ for the 1$\sigma$ scan are 
shown in Fig.~\ref{fig7} and Fig.~\ref{fig8}. These results are 
also seed averaged. As we see from the figures, 
the impact of the hadronic uncertainties are very small compared
to the CKM uncertainties in this case. The actual shape and range of
$\Delta S_{f_0K_S}$ is driven mainly by the CKM parameters.
Since $|d_{f_0\,K_S}|$ is small, the second term in
the numerator of Eq.~(\ref{dSS}) is small compared to the first term ---
$\Delta S_{f_0K_S}$ is mainly driven by the first term in the numerator.
Both terms do tend to zero, however, as $\gamma$ becomes small. 
Thus we expect to get very small values of $\Delta S_{f_0K_S}$ for
sufficiently small values of $\gamma$.
As we can see from the histograms, variations in $\gamma$ and, indeed, 
the CKM parameters impact the likely values of $\Delta S_{f_0K_S}$. 
Since $\gamma$ is never negative and the $|d_{f_0\,K_S}|^2$ term
is negligible, $\Delta S$ can only be negative if ${\rm Re}\, d_{f_0\,K_S} < 0$, 
which is an exceptionally rare occurrence.

\begin{figure}[htbp!]
\includegraphics{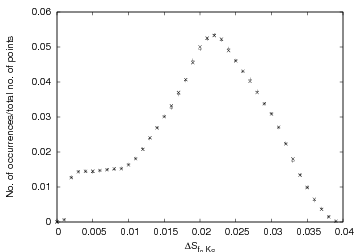}
\caption{Histogram for $\Delta S_{f_0K_S}$ varying only the CKM
parameters over a 3$\sigma$ variation. 
\label{fig9}}
\end{figure}

\begin{figure}[htbp!]
\includegraphics{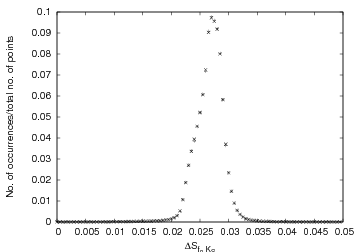}
\caption{Histogram for $\Delta S_{f_0K_S}$ varying only the hadronic
parameters over a 3$\sigma$ variation. More points exist beyond the 
window chosen for the illustration. 
\label{fig10}}
\end{figure}

When we repeat this analysis for a scan in which the 
parameters are allowed to range over 3$\sigma$, 
as shown in
Fig.~\ref{fig9} and Fig.~\ref{fig10}, we find that 
the hadronic uncertainties dominate over the CKM uncertainties.
This emerges, despite the detailed shapes in Fig.~\ref{fig9} and
Fig.~\ref{fig10}, because we use the extremal values found in 
the simulation to define the range. 
The range of $\Delta S_{f_0\,K_S}$ is thus found to be $[-0.34, 0.25]$ for the
hadronic uncertainties and $[0.00076, 0.039]$ for the CKM uncertainties. 
However, after applying the branching
ratio constraint the $\Delta S_{f_0\,K_S}$ range is found to be small, giving
$[-0.018, 0.048]$ for the hadronic uncertainties and $[0.00080, 0.038]$
for the CKM uncertainties. In the 3$\sigma$ 
scan the range of $\Delta S_{f_0\,K_S}$ is large, so that $|d_{f_0\,K_S}|$ is
no longer small, for certain values of the parameter set. It is 
the small value of the $P^c$ amplitude which is responsible for
these large values of $|d_{f_0\,K_S}|$. For $\mu = m_b$, the ranges of $\Delta S_{f_0\,K_S}$
are quite similar giving 
$[-0.0073, 0.043]$ for the hadronic uncertainties and 
$[0.00075, 0.039]$ for the CKM uncertainties 
once the branching
ratio constraint is imposed.

We have also studied the impact of
the individual hadronic parameters on the value of $\Delta S_{f_0K_S}$
for the $B \to f_0K_S$ decay process. To do this we set all the other
input parameters to their default values except the one which we want to
study. Note that we include the variations in the power corrections as
well throughout. 
We also use the same seed used to generate Figs.~\ref{fig1}
and ~\ref{fig2}. 
We find that for any hadronic input with values in the 1$\sigma$
range, the range of $\Delta S_{f_0K_S}$ is very small.
For the input parameters with values within the 3$\sigma$ range,
the largest excursions in $\Delta S_{f_0\,K_S}$ come from the first inverse moment of the 
$B$ meson 
distribution amplitude $\lambda_B$, which enters the $B\to f_0K_S$
decay amplitude through the hard spectator interaction terms. However, 
after applying the branching ratio constraint the $\Delta S_{f_0K_S}$ 
range is found to be small giving $[0.026, 0.034]$ if we employ a
renormalization scale of $\mu = m_b/2$ and giving 
$[0.026, 0.034]$ for $\mu
= m_b$. The negative
values of $\Delta S_{f_0K_S}$ found in the scan with inputs which 
range up to 3$\sigma$ of their central values are 
driven by the uncertainties 
associated with $\lambda_B$, $X_A$, and $X_H$. To see the impact of these
parameters alone on $\Delta S_{f_0\,K_S}$, 
we set all the other input parameters to
their default values and vary $\lambda_B$, $X_A$, and $X_H$ within
3$\sigma$ from their default values. The range of $\Delta S_{f_0\,K_S}$ is quite
large giving $[-0.25, 0.19]$. However, we find $\Delta
S_{f_0\,K_S}$ to be small and positive 
once we impose the branching ratio constraint, giving the range 
$[0.024, 0.036]$ and $[0.025, 0.036]$ for the two different
renormalization scales we employ, $\mu=m_b/2$ and $\mu= m_b$, respectively.
Fixing the values of 
$\lambda_B$, $X_A$, and $X_H$ to their default values, we perform a random
scan over the other input parameters within 3$\sigma$ and find
the range of $\Delta S_{f_0K_S}$ to be $[-0.0015, 0.047]$. Once we
impose the branching ratio constraint we find the range of $\Delta
S_{f_0K_S}$ to be $[0.00054, 0.046]$ and $[0.00037, 0.046]$ for $\mu
= m_b/2$ and $\mu=m_b$, respectively. 
It is interesting to note that the
range of $\Delta S_{f_0K_S}$ we find by fixing the values of
$\lambda_B$, $X_A$, and $X_H$ to their default values, and imposed
the branching ratio constraints, captures most of 
the range we find once we perform the scan with all the parameters within
3$\sigma$. However, we note that the extension of those scan results into 
negative values come from the variations associated with 
$\lambda_B$, $X_A$, and $X_H$ as well. 

In principle, $\Delta S_{f_0\,K_S}$ can have large excursions either by having a small
$P^c$ amplitude or a large $P^u$ amplitude. The first possibility can
be controlled by demanding that the theoretical model confront 
the empirical branching ratio successfully. 
As we have already noted, such small $P^c$ amplitudes 
give theoretical branching ratios which are much too small, so that 
they can be excluded by imposing the experimental branching 
ratio as a constraint. We now wish to consider whether it is
possible to have a large $P^u$ amplitude without enhancing $P^c$
as well. If it were possible, one could have a large value of
$\Delta S_{f_0\,K_S}$ without necessarily violating the branching ratio constraint. 
In the case of the default parameter set, the $a_4^q$, $a_6^q$, 
and $a_{10}^q$ --- $a_8^q$ is so small as to play no role --- 
terms interfere destructively giving small $A_{u,c}^n$ amplitudes. 
In this case, the hard spectator terms, which enter 
through the $a_4^q$ and $a_{10}^q$ terms, are similar in 
magnitude to the vertex terms in $a_6^q$. 
However, once we do a random scan over the complete 
theoretical parameter space, 
the hard spectator
terms play a dominant role in enhancing the $a_4^q$ and $a_{10}^q$ terms, 
which, in turn, enhance the $A_{u,c}^n$ amplitudes. 
The real and imaginary parts of
$a_4^q - 0.5a_{10}^q$, noting Eq.~\ref{eq:A}, e.g., can
both be very large, so that the cancellation of these terms
with $a_6^q$ is no
longer possible in these regions of parameter space. 
However, these effects act to enhance both $P^u$ and $P^c$
at the same time --- it is not possible to have a large $P^u$ 
amplitude without having a large $P^c$ amplitude as well. 
Since both $A_u^n$ and $A_c^n$ amplitudes are large, 
$\Delta S_{f_0\,K_S}$ is not enhanced. 
Thus large excursions in the hard
scattering terms can not produce large excursions in the value of 
$\Delta S_{f_0\,K_S}$. 

Our primary motivation in conducting our current analysis is to 
determine whether it is possible to control the range of 
$\Delta S_{f_0\,K_S}$ 
irrespective of the structure of the $f_0$ resonance. 
In Sec.II we noted that the structure of the $f_0$ impacts both the $B\to f_0$
form factor and $f_0$ decay constant. Thus the impact of these 
parameters alone on $\Delta S_{f_0\,K_S}$ gives us insight as to whether we
can constrain the range of $\Delta S_{f_0\,K_S}$ to small values no matter
the structure of the $f_0$. We perform a random scan varying only these
two parameters within 1$\sigma$ and 3$\sigma$ from their central values, 
both with and without the branching ratio constraint, for the particular
seed used to generate Figs.~\ref{fig1} and \ref{fig2}. 
Without the branching ratio constraint, 
the range of $\Delta S_{f_0\,K_S}$
is $[0.026, 0.028]$ for
1$\sigma$ and $[0.025, 0.028]$ for 3$\sigma$, 
respectively. Once we impose the branching ratio 
within 3$\sigma$ of the experimental value as a constraint, 
we find the range to be
$[0.025, 0.028]$ for the scan in which the $B\to f_0$ form factor and $f_0$
decay constant are varied within 3$\sigma$ of the central value. 
For $\mu = m_b$, the range of $\Delta S_{f_0\,K_S}$ in this case is also very small,
 giving $[0.026, 0.028]$. 
It is evident from our analysis that the impact of the $B\to f_0$
form factor and $f_0$ decay constant on $\Delta S_{f_0\,K_S}$ is small irrespective
of the branching ratio constraint. Apparently the strange quark content
of the $f_0$ is so important in determining $\Delta S_{f_0\,K_S}$ 
that the particular value of the ratio of the $B\to f_0$ form factor and 
$f_0$ decay constant is quite unimportant.

\section{Conclusions}
\label{cn}
In this paper, we have analyzed the two-body decay of the 
$B$ meson to one scalar meson and 
one pseudo-scalar meson, namely to the $f_0(980) K_S$ final state. 
The empirical study of this mode 
measures $\sin(2\beta)$ 
through a decay process controlled by the $b\to s$ penquin amplitude. 
This is important for several reasons. First, the study of such
modes is complementary to the study of the $b\to s$ transition in 
$B_s$ mixing. The hint that the empirical 
weak phase determined from $B_s$ mixing is larger 
than the SM prediction~\cite{Bona}, if borne out, would require new sources of
CP violation beyond that contained in the CKM paradigm. 
This makes the confirmation of, or constraints on, such effects 
in the $b\to s$ penguin modes crucial. Generally, we would like to 
assess the possible deviation of $S_f(t)$, the time-dependent CP asymmetry, 
from SM physics in a robust way, so that we can determine what new physics, 
if any, exists in these decay modes. 
To that end, the $f_0(980) K_S$ final state itself is of intrinsic interest. 
The time-dependent CP asymmetry in this case 
has no color-suppressed tree contribution; it is intrinsically
subject to less theoretical uncertainty than modes in which
such contributions are present. Moreover, its empirical study
can be complementary to that of another theoretically clean process:
$B\to\phi K_s$, as both modes occupy the $B\to KKK_s$ Dalitz plot. 
Indeed, $B\to f_0 K_s$ and $B\to\phi K_s$, assuming the $\phi$
to be ideally mixed, are the only known $b\to s$ penguin modes which 
have no color-suppressed tree contributions.

We have investigated the size of $\Delta S_{f_0\,K_S}$, i.e., the
deviation of $S_f(t)$ from $\sin(2\beta)$, in 
the $B\to f_0K_S$ decay process in the SM using the QCD factorization 
approach, assuming the $f_0$ to be a $q\bar{q}$ state. 
We employ a parameter scan to probe a broad range of 
possible theoretical models, exploring variations in the inputs at
the 3$\sigma$ level and ill-known ${\cal O}(\Lambda_{QCD}/M_B)$
corrections with 100\% uncertainty.  
The $B\to f_0K_S$ decay mode has been 
studied by other authors within the 
QCD factorization approach~\cite{ck1,cheng,soni}. 
Our calculation differs most significantly from 
this earlier work in its treatment of the $B\to f_0$ 
form factor, for which, for concreteness, we employ
the CQM approach of Refs.~\cite{Gatto,Deandrea1,Deandrea2,Polosa}. 
The earlier work simply asserts that the $B\to f_0$ 
form factor ought be comparable to that of $B\to\pi_0$. 
Our numerical studies support this, though we assign
a large uncertainty to our assessment. The parameter
scan technique we employ borrows heavily from the
work of Beneke~\cite{beneke}, in which the deviations
to $S_f(t)$ from $\sin(2\beta)$ were studied for
the $B\to (\pi^0 \rho^0, \eta, \eta^\prime, \phi) K_s$ decay
modes. That work eschewed the $B\to f_0 K_s$ decay mode due to the 
uncertain $q\bar q$ structure of the $f_0$~\cite{MBcom}; 
thus we have studied the possible numerical uncertainty 
incurred by this with great care. Our work also differs from Beneke's
in that we explore the theoretical model space for 
input parameters which vary within 3$\sigma$ of their default
values, as well as identify the parameter 
morphologies which give the largest excursions in $\Delta S_{f_0\,K_S}$. 

The assumed quark structure of the $f_0(980)$ does enter
in the evaluation of the hadronic matrix elements
and in the assessment of the $f_0$ decay constant and the 
$B\to f_0$ form factor in particular. 
We have investigated the range of $\Delta S_{f_0\,K_S}$ which
results from 
varying either the $B\to f_0$ form factor and the $f_0$ scalar decay
constant, after imposing the branching ratio constraint, 
giving $[0.026, 0.028]$ for 
variations within 1$\sigma$ and 
$[0.025, 0.028]$ for variations within 3$\sigma$. 
The value of $\Delta S_{f_0\,K_S}$ is simply not 
sensitive to the value of the $F^{Bf_0}$ form factor
and the $f_0$ decay constant.
Although we can accommodate the experimental branching ratio 
in a $q\bar q$ model of the $f_0$, we can not draw conclusions 
concerning the structure of the $f_0$ on the basis of our analysis.

Let us summarize our results for the range of $\Delta S_{f_0\,K_S}$ and $C_{f_0\,K_S}$, 
which emerge from a scan of the complete theoretical parameter space, 
allowing for uncertainties in the infrared-cutoff
dependent ${\cal O}(\Lambda_{QCD}/M_B)$ corrections, 
which we characterize by the parameters $X_A$ and $X_H$. 
In limited regions of the parameter space, the $P^c$ amplitude can become
small, driving large excursions in $\Delta S_{f_0\,K_S}$. Such excursions
are removed, however, by demanding that the theoretical model 
confront the empirical branching ratio up to some tolerance,
that is, up to 1$\sigma$ for the 1$\sigma$ scans and up
to 3$\sigma$ for the 3$\sigma$ scans.  
Once we demand that our theoretical models satisfy 
the branching ratio constraint, the possible seed averaged range in
$\Delta S_{f_0 K_S}$ is greatly reduced, giving 
$[0.018,0.033]$ and 
$[-0.019, 0.064]$ for the scans in which the
parameters are allowed to range within 1$\sigma$ and 3$\sigma$ 
of their central values, respectively. In comparison, 
we find that $C_{f_0\,K_S}$ ranges over $[-0.013, -0.0024]$ 
and $[-0.045, 0.051]$, respectively. Generally, we find
$d_{f_0\,K_S}$ to be sufficiently small that the values of  
$\Delta S_{f_0\,K_S}$ and $C_{f_0\,K_S}$
are uncorrelated, note Fig.~\ref{fig1}. It is worth emphasizing
that we estimate the ranges using the extremal values of our
parameter scans. This is a conservative approach. If we were
to define the range of $\Delta S_{f_0\,K_S}$ to capture, 
e.g., 95\% of the models 
about its most likely value in theory space, 
we would have much smaller ranges. 
Retaining our current approach, we could also sharpen our 
estimates once improved measurements of the branching
ratio become available. 

Nevertheless, let us proceed to compare our range for
$\Delta S_{f_0\,K_S}$ with the empirical result. 
The time-dependent CP asymmetry 
induced by the interference between $B\bar B$ mixing
and direct decay, as well as the direct CP
asymmetry, measured by Belle and BaBar~\cite{talk,Aubert1, Abe1}
are $0.85\pm0.07$
and $0.08\pm0.12$, respectively. The value of $\sin(2\beta)$ measured
by Belle and BaBar~\cite{talk} in $B \to J/\Psi K_S$ decays and
related charmonium modes is
$0.668\pm0.026$, so that the empirical value of $\Delta S_{f_0\,K_S}$ is 
$0.18\pm0.10$ for the $B \to f_0 K_S$ decay mode. 
Its error is substantially larger than our largest estimated
value of $\Delta S_{f_0\,K_S}$, so that further refinement of the 
experimental results is warranted. If we compare our results
with those of Ref.~\cite{beneke} for 
the $B\to (\pi^0 \rho^0, \eta, \eta^\prime, \phi) K_s$ decay
modes, we see that the $B\to f_0 K_S$ mode compares favorably: 
it yields small values of $\Delta S_f$, as do the 
$B\to \phi K_s$ and $B\to \eta^\prime K_s$ modes. 
Since both $B\to f_0 K_s$ and $B\to\phi K_s$ decays lack 
color-suppressed tree contributions, the estimation of $\Delta S_f$
should be particularly reliable, so that 
significant differences 
in the empirical values of $\Delta S_f$ in these modes could truly 
signal new physics. 

We find that the largest single variation in our predictions of the 
decay amplitudes 
come from the uncertainty
associated with the inverse moment of $B$ meson distribution amplitude
$\lambda_B$ which enters the decay amplitude through the hard
scattering terms. In our scans for which the parameters
range over 3$\sigma$ of their central values, 
the hard scattering terms can dominate
over all other terms in the decay amplitude and produce large values
of the $A_{u,c}^n$ amplitudes. 
The cancellation of $a_4^q$ with $a_6^q$
reflected
in the default values of the parameter set is no longer possible
for such large values of the hard scattering amplitude. This mechanism 
generalizes to any other mode where $a_4^q$ and $a_6^q$ interfere
destructively. However, these large excursions in the $A_{u,c}^n$
amplitudes do not generate large excursions in $\Delta S_{f_0\,K_S}$, as they
also act to enhance the charm penguin amplitude. Nevertheless,
the extremal values of $\Delta S_{f_0\,K_S}$, and its negative values
in particular, 
do come from the uncertainties 
associated with $\lambda_B$, $X_A$, and $X_H$. One expects
this observation to be relevant to other $b\to s$ mode as well, 
though 
if there is a color-suppressed tree amplitude, there is more freedom
for $\Delta S$ to be large. 

In future studies we would like to consider how 
the finite width of the $f_0(980)$ resonance can impact 
the value of $\Delta S_{f_0 K_S}$ determined from $B\to f_0 K_S \to
\pi^+\pi^-K_S$ and $B\to f_0 K_S \to K^+K^-K_S$ decay processes.
In particular, we wish to consider the role of the 
$f_0\to \pi^+\pi^-$ and $f_0\to K^+K^-$ scalar
form factors,  
as well as that of possible OZI-violating effects, in the
determination of $\Delta S_{f_0 K_S}$ 
in a later
publication~\cite{etal}.
\vskip 1.0cm
\acknowledgments
We thank Timo A. L{\"a}hde, Ulf-G. Mei{\ss}ner, 
and Jos{\'e} Oller for 
discussions concerning 
the scalar form factor in the $f_0(980)$ mass region 
and Martin Beneke for a discussion concerning the modelling
of power corrections within QCD factorization. We also thank A. D. Polosa for
useful discussions regarding the CQM model and the determination of
the various form factors within this model.
We thank 
the Institute for
Nuclear Theory (INT) 
for gracious hospitality and acknowledge partial support from the 
U.S. Department of Energy under 
contract DE--FG02--96ER40989. S.G. 
thanks the SLAC theory group for gracious hospitality as well, and 
R.D. acknowledges 
a Huffaker Travel Scholarship from the University of Kentucky 
in generous support of his visit to the INT. S.G. also acknowledges
partial support from the National Science Foundation under 
Grant No. NSF PHY05-51164 at its completion.

\appendix
\section{Annihilation Amplitudes in QCD Factorization}
\label{app}
The weak annihilation contributions $b_3$ and $b_{3,EW}$ in the $B\to
f_0K_S$ decay processes are expressed as linear combinations of the
annihilation amplitudes $A^{i,f}_{1,2,3}$. Explicit expressions for 
these amplitudes in terms of the light-cone distribution amplitudes for
scalar and pseudoscalar mesons are given in \cite{bbns,cheng}. 
The general expressions for the 
leading-twist, light-cone distribution amplitudes for 
pseudoscalar and scalar mesons are written as
\begin{eqnarray}
&& \Phi_P(x,\mu) = 6\,x\,(1-x)\Bigg\{1 +
\sum_{n=1}^{\infty}\,\alpha_n^P(\mu)\,C_n^{3/2}\,(2\,x - 1)\Bigg\} \nonumber
\\
&& \Phi_S(x,\mu) = \bar{f}_S\,6\,x\,(1-x)\Bigg\{B_0 +
\sum_{m=1}^{\infty}\,B_m(\mu)\,C_m^{3/2}\,(2\,x - 1)\Bigg\} \,.
\label{gegen}
\end{eqnarray}
Here $B_0$ is zero, as are all the even Gegenbauer moments, in the SU(3)
limit. In order to present simple expressions for the
$A^{i,f}_{1,2,3}$ amplitudes, we truncate the expansion
of the pseudoscalar meson after the second Gegenbauer polynomial and 
that of the scalar meson after the third Gegenbauer polynomial.

We perform the integration over the 
light-cone distribution amplitudes in such a way that we always
have $I_{xy} = I_{yx}$, where
\begin{eqnarray}
I_{xy} = \int_{0}^{1-\epsilon}\,dx\int_{\epsilon}^{1}\,dy\,f(x,y) \,,
\nonumber \\
I_{xy} = \int_{\epsilon}^{1}\,dy\int_{0}^{1-\epsilon}\,dx\,f(x,y)\,. 
\end{eqnarray} 
We retain finite $\epsilon$ throughout the calculation and put
$\epsilon\to 0$ only at the end to avoid dropping finite contributions.
This amounts to a model of the power corrections which
differs slightly from that employed previously~\cite{BN,cheng}, giving
slightly different results for $A_3^i(PS)$ and $A_3^i(SP)$. Instead of
the $\pi^2/3$ found in Refs.~\cite{BN,cheng}, we get $\pi^2/6$.  That is, 
the annihilation amplitudes can be written as
\begin{eqnarray}
A_1^i(PS) &=& \pi \alpha_s f_K\bar
f_{f_0}^{s}\Big[B_1[(180-18\pi^2)+\alpha_1^K(3726-378\pi^2)+\alpha_2^K(27720-280
8\pi^2)]
\nonumber \\
&&+
B_3[(593-60\pi^2)+\alpha_1^K(37305-3780\pi^2)+\alpha_2^K(714168-72360\pi^2)]
\nonumber \\
&&+
(3\alpha_2^K-3\alpha_1^K+3)[B_1(18X_A-36)+B_3(60X_A-190)]-2\bar
r_{\chi}^Sr_{\chi}^K X_A^2\Big] \, ,
\end{eqnarray}
\begin{eqnarray}
A_1^i(SP) &=& \pi \alpha_s f_K\bar
f_{f_0}^{u}\Big[B_1[(-540+54\pi^2)+\alpha_1^K(3726-378\pi^2)+\alpha_2^K(-13860+1
404\pi^2)]
\nonumber \\
&&+
B_3[(-5930+600\pi^2)+\alpha_1^K(124350-12600\pi^2)+\alpha_2^K(-1190280+120600\pi
^2)]
\nonumber \\
&&+
(-3B_3-3B_1)[6(X_A-1)+18\alpha_1^K(X_A-2)+12\alpha_2^K(3X_A-8)]+2\bar
r_{\chi}^S
r_{\chi}^K
X_A^2\Big] \,,
\end{eqnarray}
\begin{eqnarray}
A_3^i(PS) &=& \pi \alpha_s f_K\bar
f_{f_0}^{s}\Big[r_{\chi}^K[18B_1(X_A^2-4X_A+4+\frac{\pi^2}{6})+60B_3(X_A^2-\frac
{19}{3}X_A+\frac{191}{18}+\frac{\pi^2}{6})]
\nonumber \\
&&+
\bar
r_{\chi}^S[6(X_A^2-2X_A+\frac{\pi^2}{6})-18\alpha_1^K(X_A^2-4X_A+4+\frac{\pi^2}{
6})
\nonumber \\
&&+
36\alpha_2^K(X_A^2-\frac{16}{3}X_A+\frac{15}{2}+\frac{\pi^2}{6})]\Big]\,,
\end{eqnarray}
\begin{eqnarray}
A_3^i(SP) &=& \pi \alpha_s f_K\bar f_{f_0}^{u}\Big[r_{\chi}^K[-18B_1(X_A^2-4X_A+4+\frac{\pi^2}{6})-60B_3(X_A^2-\frac{19}{3}X_A+\frac{191}{18}+\frac{\pi^2}{6})]
\nonumber \\
&&-
\bar
r_{\chi}^S[6(X_A^2-2X_A+\frac{\pi^2}{6})+18\alpha_1^K(X_A^2-4X_A+4+\frac{\pi^2}{
6})
\nonumber \\
&&+
36\alpha_2^K(X_A^2-\frac{16}{3}X_A+\frac{15}{2}+\frac{\pi^2}{6})]\Big]\,,
\end{eqnarray}
\begin{eqnarray}
A_3^f(PS) &=& \pi \alpha_s f_K\bar
f_{f_0}^{s}X_A\Big[r_{\chi}^K[6B_1(6X_A-11)+B_3(120X_A-374)]
\nonumber \\
&&-
\bar
r_{\chi}^S[6(2X_A-1)-6\alpha_1^K(6X_A-11)+\alpha_2^K(72X_A-186)]\Big]\,,
\end{eqnarray}

\begin{eqnarray}
A_3^f(SP) &=& \pi \alpha_s f_K\bar
f_{f_0}^{u}X_A\Big[r_{\chi}^K[6B_1(6X_A-11)+B_3(120X_A-374)]
\nonumber \\
&&-
\bar
r_{\chi}^S[6(2X_A-1)+6\alpha_1^K(6X_A-11)+\alpha_2^K(72X_A-186)]\Big] \,.
\end{eqnarray}
\pagebreak


\end{document}